# Near-complete protein structural modelling of the minimal genome

Joe G Greener*, Nikita Desai*, Shaun M Kandathil, David T Jones[+]

Department of Computer Science, University College London, Gower Street, London WC1E 6BT, UK
* These authors contributed equally
[+] To whom correspondence should be addressed (email: d.t.jones@ucl.ac.uk)

## Abstract

Protein tertiary structure prediction has improved dramatically in recent years. A considerable fraction of various proteomes can be modelled in the absence of structural templates. We ask whether our DMPfold method can model all the proteins without templates in the JCVI-syn3.0 minimal genome, which contains 438 proteins. We find that a useful tertiary structure annotation can be provided for all but 10 proteins. The models may help annotate function in cases where it is unknown, and provide coverage for 29 predicted protein-protein interactions which lacked monomer models. We also show that DMPfold performs well on proteins with structures released since initial publication. It is likely that the minimal genome will have complete structural coverage within a few years.

## Introduction

The last few years have seen vast improvements in the prediction of protein tertiary structure from amino acid sequence [1–5]. The combination of residue-residue covariation information from sequence alignments [6] and deep learning techniques [7] mean that accurate models can be obtained for sequences in the absence of structural templates. As performance improves to the point where more than two thirds of targets without templates were predicted correctly by AlphaFold in the Critical Assessment of protein Structure Prediction (CASP) 13 experiment [8], the answer to the question of how much of the proteome we can structurally annotate changes.

There have been attempts to apply *de novo* protein structure prediction to the regions of known sequence space without templates [9], including PConsFold2 [10,11] and a study from the Baker group using metagenomic data [12]. Recently, proteome-scale prediction has also been applied to predicting the structures of protein complexes [13]. In our paper presenting DMPfold, which makes many of the recent advances available to the community, we used deep learning-based prediction of structural constraints to provide high-confidence models for 25% of Pfam families without templates. We show that this improves coverage of various model organism proteomes [14], with such approaches being important for systems-level insights [15]. However, none of the model organisms we studied had complete or even near-complete coverage.

Studies to discover essential genes have a rich history [16,17] but determining the minimal set of genes that leads to survival and replication is complicated by multiple proteins, each non-essential, carrying out an essential function. The minimal set of genes is also dependent on the environment the organism is in, with many non-essential genes providing adaptability to diverse environmental conditions [18]. The discovery of the synthetic genome JCVI-syn3.0 using genome transplantation and global transposon mutagenesis provides the smallest genome to date, consisting of 438 protein-coding genes and 35 RNA-coding genes [18]. The original authors assigned function to all but 149 genes; subsequent studies have tackled this problem [19–21] with one study using a variety of *in silico* methods to assign function to 66 of the remaining 149 [21].

In this study we investigate whether protein structure prediction has advanced to the level where the whole JCVI-syn3.0 proteome can be modelled. Most of the 438 proteins have templates available in the PDB; we use DMPfold and metagenomic data to provide models for nearly all of the remainder, leaving only 10 without tertiary annotation. These models will be useful to further studies predicting functions for the proteins, provide monomer models to predict protein-protein interactions, and point to a future where reliable genome-scale structure annotation is possible.

## Results

**Protein structural modelling of the minimal genome**

We find that 389 of 438 (89%) protein sequences in the minimal genome have a template available in the PDB by searching with HHsearch [22] and GenTHREADER [23]. This high fraction likely represents the bias of structural determination studies towards essential and widespread proteins. The 49 remaining sequences range from 46 to 978 residues in length and represent single and multi-domain proteins. See Figure 1 for the eventual classification of these proteins according to this study and Figure 2 for the collection of models. They are not generally predicted to have disordered regions, with only 2 having more than 20% of residues predicted disordered by DISOPRED3 [24]. 24 of the 49 are predicted to be transmembrane proteins [21], which at 49% is higher than the 21% of the 438 proteins predicted to be transmembrane. This likely represents the scarcity of templates available for transmembrane proteins.

For the 49 sequences we obtained related sequences, including from metagenomic databases, and ran our DMPfold method with default parameters. We also split sequences longer than 300 residues into domains using DomPred [25] and visual inspection of initial full-chain DMPfold models. We find that 15 proteins have a full-length or domain model (12 full-length, 3 domain) with a predicted TM-score [26] to the native structure of at least 0.5; according to the cross-validation in the DMPfold study, this indicates an 83% chance of the model having an actual TM-score of at least 0.5 [14]. A TM-score of 0.5 generally means the model has the correct fold. We call this set the "high-confidence" set. Another 10 proteins (3 full-length, 7 domain) have models with predicted TM-score of at least 0.35, indicating a 61% chance of the model having an actual TM-score of at least 0.5. We call this set the "medium-confidence" set.

We scan the full-length DMPfold models of the remainder against the PDB using mTM-align [27] and find that 10 have a match in the PDB with TM-score of at least 0.5. It is generally accepted that when a model matches a known fold like this it is likely to be correct, provided that profile drift has not occurred in the alignment. We call this set the "matches known structure" set. By examining the PSIPRED secondary structure predictions of the remaining 14 sequences it can be seen that 4 of the sequences (sequence lengths 46, 60, 86 and 145) have two strongly-predicted helices and no other predicted secondary structure. These sequences likely adopt a simple topology with two helices joined by a loop. One of these sequences, MMSYN1_0479, has a block of 20 residues between the helices that is predicted disordered. Counting this "simple topology set" as a tertiary structure annotation, 428 of the 438 minimal genome protein sequences are given a tertiary structure annotation. Models for the remaining 10 sequences may still be correct or partially correct. Details for these sequences are given in Table 1. 3 predicted domains on 3 of the modelled 428 sequences do not have a tertiary structure annotation but are on a sequence that does have at least one domain with an annotation. All models and intermediate data are made available at http://bioinf.cs.ucl.ac.uk/downloads/minimal_genome/mg_structure_data_v1.tar.gz.

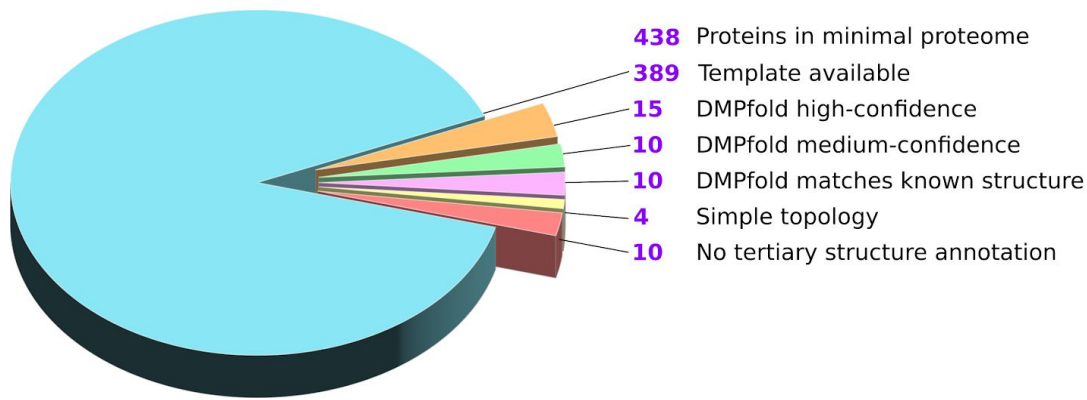

**Figure 1:** Structural modelling of the minimal proteome. The number of proteins in each category is shown. Models for the 49 proteins with no templates available are made available in this study. Some of the DMPfold high-confidence and medium-confidence models are for domains rather than the full-length sequence, meaning that there are a further 3 predicted domains that do not have a tertiary structure annotation but are on a sequence that does have at least one domain with an annotation.

**A** DMPfold high-confidence

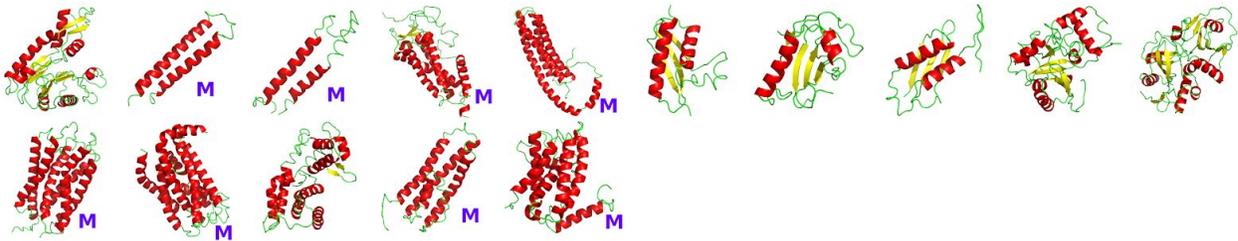

**B** DMPfold medium-confidence

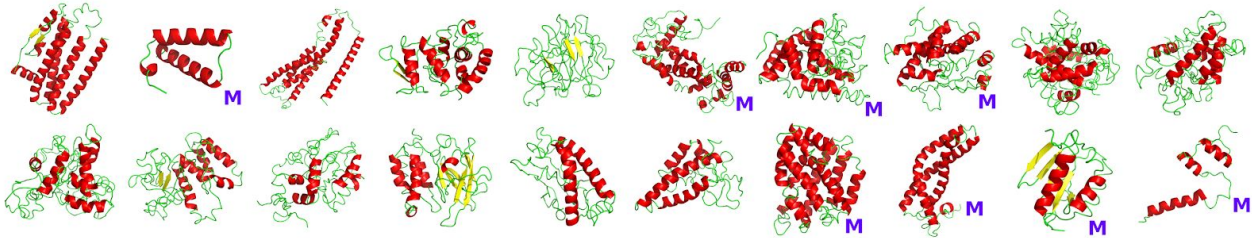

**C** DMPfold matches known structure

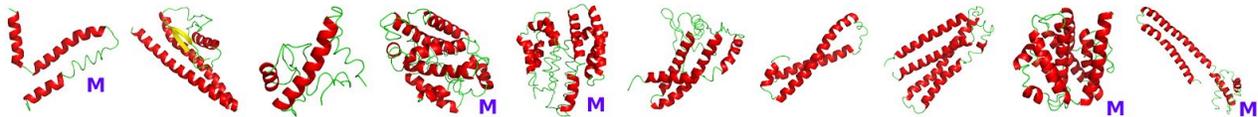

**D** Simple topology

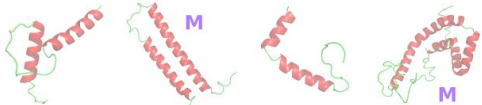

**E** No tertiary structure annotation

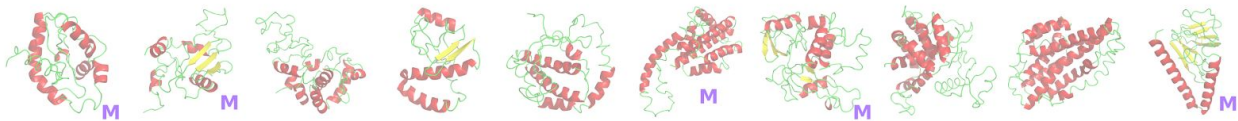

**Figure 2:** Models for proteins in the minimal proteome without structural templates. Models are shown coloured by secondary structure for each category. Some of the DMPfold high-confidence and medium-confidence models are for domains rather than the full-length sequence. Models in the two categories judged to be low-confidence, simple topology and no tertiary structure annotation, are shown with partial transparency. Some of these models may still be correct or partially correct. Models belonging to a protein predicted to be transmembrane according to [21] are marked with a purple M.

| Gene name | Sequence length | Alignment depth | Alignment $N_{eff}$ | Fraction disordered by DISOPRED3 | PSIPRED prediction | Predicted trans-membrane | Function category annotation |
|---|---|---|---|---|---|---|---|
| MMSYN1_0248 | 192 | 2461 | 863 | 0.08 | mainly beta | yes | Transport |
| MMSYN1_0285 | 268 | 117 | 30 | 0.00 | alpha | no | Translation |
| MMSYN1_0330 | 275 | 112 | 25 | 0.09 | mainly alpha | no | Nucleotide salvage |
| MMSYN1_0332 | 239 | 1416 | 474 | 0.08 | alpha/beta | yes | Transport |
| MMSYN1_0345 | 231 | 106 | 27 | 0.07 | alpha | yes | Cofactor transport and salvage |
| MMSYN1_0387 | 205 | 261 | 82 | 0.25 | alpha/beta | no | tRNA modification |
| MMSYN1_0503 | 133 | 126 | 45 | 0.13 | alpha/beta | no | Unclear |
| MMSYN1_0511 | 207 | 22 | 3 | 0.06 | mainly alpha | no | Unclear |
| MMSYN1_0632 | 158 | 588 | 286 | 0.15 | alpha/beta | yes | Transport |
| MMSYN1_0777 | 161 | 80 | 30 | 0.04 | alpha/beta | yes | Unclear |

**Table 1:** Details of the 10 proteins in the minimal genome which are not given a confident structural model by this study. $N_{eff}$ is the number of clusters returned by CD-HIT [28] at a sequence identity threshold of 62%. "Predicted transmembrane" and "Function category annotation" are taken from [21]. Low confidence models of these proteins are shown in Figure 2E.

**Predicting function**

The models provided in this study can assist in predicting function for proteins in the minimal genome. A previous study that assigns function [21] made use of template-based models from the Phyre2 server [29]. *De novo* structure prediction extends these techniques to proteins without templates but confident *de novo* models. In addition, our structure predictions offer corroborating evidence for current function annotations. MMSYN1_0375 has no function annotation but the DMPfold model has structural similarity to many proteins in the PDB, with the top 10 hits all being RNase E and a closest TM-score of 0.56. In particular the model matches the region that interacts with RNA 5'-terminal monophosphate. MMSYN1_0530 has no function annotation but the DMPfold model has structural similarity to many proteins in the PDB, with 9 of the top 10 hits being sarco/endoplasmic reticulum $Ca^{2+}$-ATPase (SERCA) and a closest TM-score of 0.58. The models and PDB structures for these cases are shown in Figure 3A.

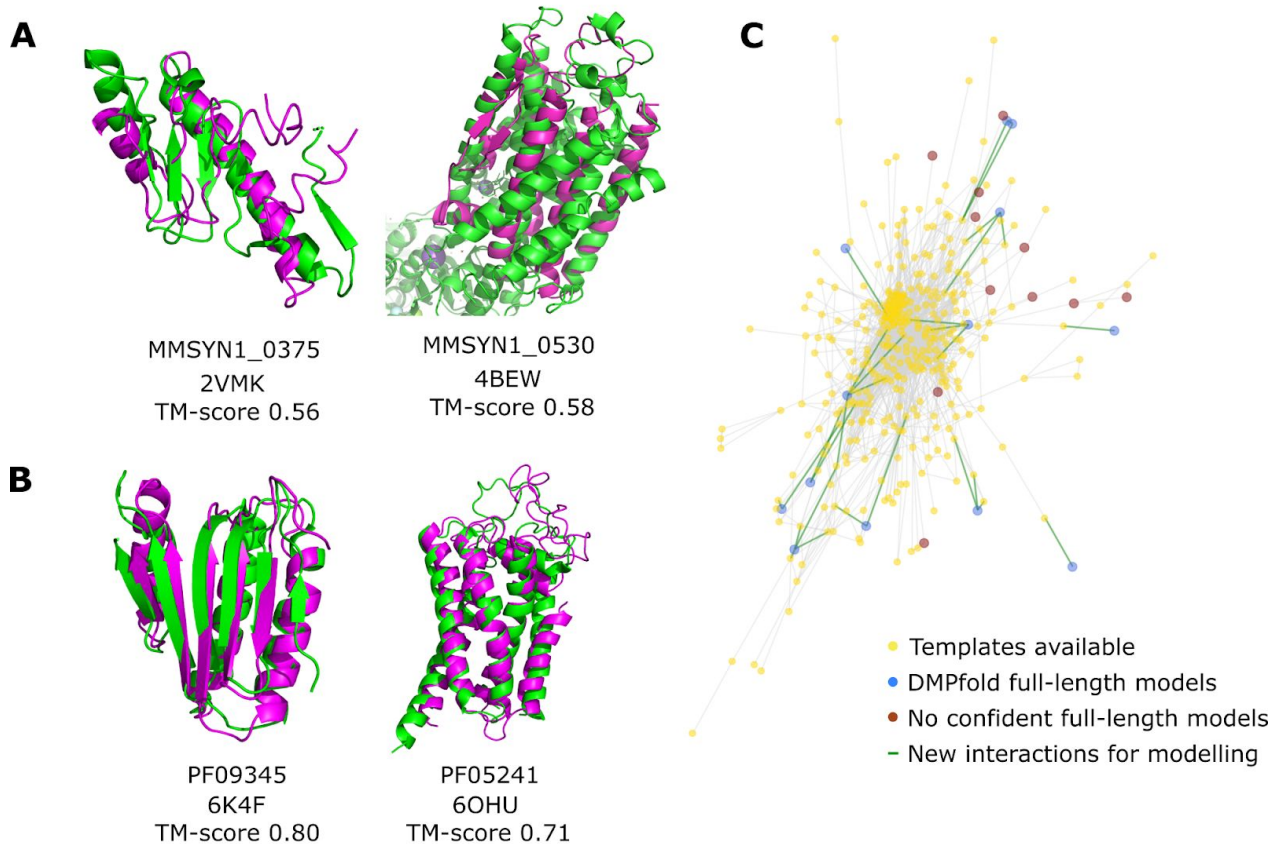

**Figure 3:** (A) Matches of models to the PDB. The DMPfold model for MMSYN1_0375, which has no functional annotation, is similar to many proteins in the PDB including RNase E. The superimposition to the RNA 5'-terminal monophosphate region of 2VMK is shown. The DMPfold model for MMSYN1_0530, which has no functional annotation, has matches to SERCA. The superimposition to 4BEW is shown. (B) Examples from the continuous validation of DMPfold. In each case the superimposition of the Pfam model from [14] to the newly-released PDB structure is shown. The TM-scores given are calculated using TM-align. (C) The protein-protein interaction network predicted from sequence by STRING [30], showing the interactions that could be modelled using monomer *de novo* models provided by this study.

**Predicting the structure of protein complexes**

Next, we ask how many additional protein complexes could be modelled using the *de novo* monomer models we have generated. To get an indication of the protein-protein interactions present in the proteome we searched the sequences in the STRING database [30], and selected the entries with highest identity (identity >40) and bitscores for each amino acid sequence. We then considered interacting proteins as those with a combined score of at least 0.7, the high-confidence threshold as defined in the STRING database. This predicts 3,499 interactions involving 389 of 438 proteins, shown in Figure 3C. 57 of these interactions involve proteins without templates, with 24 of the 49 proteins without templates being predicted to have at least one interaction. The full-length models in the high-confidence, medium confidence and matches structure categories provide monomer models for 29 of these interactions. 28 are between a *de novo* model and a template model and one is between two *de novo* models. As methods for predicting protein complexes using coevolutionary data improve, models for these interactions can be predicted [31].

**Continuous validation of DMPfold Pfam models**

In the DMPfold study we modelled Pfam families without available templates [14]. The final version of the paper reported that during the review process 9 Pfam families had structures deposited in the PDB for the first time, and our models had a TM-score of at least 0.5 to the native structure in 8 out of 9 cases. We repeated the same analysis for the period of time the paper has been published and find in this case that 11 out of the 11 models with newly-released structures have a TM-score of at least 0.5 to the native structure. Some examples are shown in Figure 3B. These results exceed the predicted success rate of 83% and indicate the accuracy of DMPfold in a blind test setting. A similar validation procedure can be applied to our minimal genome models as structures become available.

## Discussion

As shown in Table 1, the 10 proteins without tertiary structure annotation range from 133 to 275 residues in length, with 5 being predicted as transmembrane proteins. 3 have no function annotation. 8 of the 10 have fewer than 1,000 sequences in the sequence alignment, though only 2 of the 10 have fewer than 100 sequences in the alignment. Only one, MMSYN1_0387, appears to have significantly disordered regions with 25% of residues predicted disordered by DISOPRED3. There does not appear to be anything in particular that is special about these proteins, and it is likely that confident models will be produced in the future as more sequence data becomes available and protein structure prediction methods continue to improve. They could also be targeted for experimental study by groups like the Structural Genomics Consortium, or provided as targets in a future CASP or CASP_Commons experiment.

## Methods

The 438 protein sequences of protein-coding genes in JCVI-syn3.0 [18] were considered. HHsearch [22] was used to search these against the PDB70 HMM library and sequences with hits of 90% probability and 80% coverage were determined to have available templates. The remaining sequences were run through GenTHREADER [23] and hits with "certain" confidence were also determined to have available templates.

The 49 sequences without templates were taken forward for *de novo* protein structure prediction. Alignments were formed by searching for sequences from 3 sources. HHblits [32] searches were carried out against the 2018_08 version of the UniClust30 database, the BFD database [33], and our custom pipeline [34] that uses sequences from the EBI MGnify database [35]. Alignments were formed by combining hits from the 3 sources and removing duplicates. The alignments were run through DMPfold and models with predicted TM-score [26] of at least 0.5 and at least 0.35 were formed into high-confidence and medium-confidence sets respectively. The remainder were split into domains manually using DomPred [25] and visual inspection of the DMPfold models. Alignment generation was carried out again with the domain sequences and DMPfold was run again with the resulting alignments. Domain models with predicted TM-score of at least 0.5 and at least 0.35 were added to the high-confidence and medium-confidence sets respectively.

DMPfold models for sequences not having a high or medium-confidence model (whole length or domain) were searched against the PDB using mTM-align [27]. Sequences with hits of TM-score 0.5 or more were put in a set of models that match known folds. The PSIPRED [36] secondary structure predictions of the remainder were examined and cases with two strongly predicted helices and no other secondary structural elements were considered part of a set with simple topology.

Continuous validation of DMPfold models was carried out by searching modelled sequences against the PDB using blastp [37] with an E-value of $10^{-3}$.

# Data availability

All data is made available at http://bioinf.cs.ucl.ac.uk/downloads/minimal_genome/mg_structure_data_v1.tar.gz. The DMPfold method is described in [14] and available at https://github.com/psipred/DMPfold.

# Acknowledgements

We thank members of the group for valuable discussions and comments. This work was supported by the European Research Council Advanced Grant "ProCovar" (project ID 695558).